\documentclass[11pt]{article}
\usepackage{emulateapj,pstricks,apjfonts}

\def\asca	{{\em ASCA}\/}
\def\rosat	{{\em ROSAT}\/}
\def\chandra 	{{\em Chandra}\/}
\def\bi{\bfseries\itshape}
\def\deg	{$^{\circ}$}

\begin{document}

\submitted{Poster presented at {\em X-ray Astronomy 2000}, Palermo,
September 4-8, 2000}

\title{TEMPERATURE STRUCTURE OF FOUR MERGING CLUSTERS OBTAINED WITH {\em
CHANDRA}}

\author{M. Markevitch, A. Vikhlinin, P. Mazzotta, L. VanSpeybroeck (SAO)}

\begin{abstract}

We present preliminary \chandra\ results on $z\approx0.2$ clusters A665,
A2163 and A2218, and a $z=0.05$ cluster A754.  For A754, A665 and A2163, we
have derived first high-resolution projected gas temperature maps. All three
show strong spatial temperature variations in the inner
$r<0.5-1\;h^{-1}_{50}$ Mpc regions, indicating ongoing mergers. The maps
reveal a probable shock in front of a moving cluster core in A665, a rather
complicated temperature distribution in the center of A2163, and possibly a
merger of three subclusters in A754. At greater off-center distances, radial
profiles for A2163 and A2218 show a temperature decline, in agreement with
earlier \asca\ results.

\end{abstract}

\section*{INTRODUCTION}

Clusters of galaxies, the biggest collapsed mass aggregations in the present
universe, are formed through the hierarchical process of gravitational
infall, collision and merging of smaller groups and clusters and a
subsequent relaxation of all the matter components in the cluster. Each
merger is a very energetic event. A large fraction of the kinetic energy of
the collision goes into shock heating of the intracluster gas (e.g.,
Schindler \& M\"uller 1993; Roettiger et al.\ 1993 and later works), a
process that has great importance for cluster physics and that we can
readily observe using the \chandra's combination of energy coverage and
superb spatial resolution. Here we present preliminary results from ACIS-I
on four hot clusters, A665, A2163, A2218, and A754. X-ray images of all
these clusters show structure of various degrees of complexity; however, in
the absence of the characteristic temperature signatures, it may be a result
of projection rather than a merger. For A754, unambiguous data existed
before that it is a merger, from both the optical and X-ray temperature data
(Zabludoff \& Zaritsky 1995; Henry \& Briel 1995; Henriksen \& Markevitch
1996). For A2163, some indication of a nonrelaxed state from the gas
temperature and optical data existed (Markevitch et al.\ 1994; Squires et
al.\ 1997); and for A2218, weak lensing mass reconstruction suggested two
mass concentrations (Kneib et al.\ 1996). For A665, there has been no
suggestion of a nonrelaxed state other than an asymmetric X-ray image.
A665, A2218 and A2163 also are famous SZ sources (e.g., Birkinshaw et al.\
1991, references therein and later works) and detailed gas temperature maps
are critical for correct interpretation of the SZ data.

\section*{DATA ANALYSIS}

\begin{figure*}[t]
\pspicture(0,5)(18.5,12)

\rput[tl]{0}(-0.3,12){\epsfxsize=6.5cm \epsfclipon
\epsffile[18 144 592 676]{A665_tprof.ps_b}}

\rput[tl]{0}(5.9,12){\epsfxsize=6.5cm \epsfclipon
\epsffile[18 144 592 676]{A2163_tprof.ps_b}}

\rput[tl]{0}(12.1,12){\epsfxsize=6.5cm \epsfclipon
\epsffile[18 144 592 676]{A2218_tprof.ps_b}}

\rput[cc]{0}(5.3,11.6){\large\bi a}
\rput[cc]{0}(11.6,11.6){\large\bi b}
\rput[cc]{0}(17.8,11.6){\large\bi c}

\rput[tl]{0}(0,6.0){
\begin{minipage}{18.5cm}
\small\parindent=3.5mm
{\sc Fig.}~1.---Projected radial temperature profiles of (a) A665, (b)
A2163, (c) A2218. The A665 and A2163 profiles are overlaid on the \asca\
results from Markevitch (1996).
\par
\end{minipage}
}
\endpspicture
\end{figure*}

A665, A2163 and A2218 were observed by ACIS-I for short exposures of 9 ks,
10 ks, and 18 ks, respectively (clean exposure after the rejection of
background flares, if any), which is sufficient for derivation of crude
temperature maps and radial temperature profiles. A754 was observed by
ACIS-I for a clean exposure of 39 ks, sufficient for a detailed temperature
map of the central $16'\times 16'$ region covered by ACIS-I.

The ACIS gain tables, quantum efficiency curves and spectral response
matrices latest as of September-October 2000 were used. The detector + sky
background was estimated using the appropriate blank sky datasets,
correcting for a slow secular decrease of the background rate (Markevitch
2000). The background uncertainty of $\pm5$\% was included in quadrature to
the statistical uncertainties of the temperature values (it is significant
for the radial profiles at high radii). Point sources were excluded from the
analysis, and the energy band of 0.8--9 keV (or close) was used. The
CTI-induced nonuniformity of the ACIS-I quantum efficiency was corrected
using an approximate model function (Vikhlinin 2000). For the A754, A665 and
A2218 observations made at the $-110$\deg C ACIS temperature, this
correction is important, while for A2163 observed later at $-120$\deg C, it
is much less significant. We also used an additional experimental
position-independent correction factor of $\approx 0.9$ for the ACIS-I
quantum efficiency below $E=2$ keV to account for the difference between
ACIS-S3 and ACIS-I observations of a calibration SNR (Vikhlinin 2000).  For
hot clusters, ignoring it results in spuriously high Galactic absorption
columns and temperatures and their dependence on the adopted lower energy
cut. Once these corrections were applied, all fits were acceptable and
independent of the energy band used. The absorption column was fixed at the
Galactic values for A754, A665 and A2218, and at the \rosat\ PSPC value for
A2163 (which is higher than the Galactic one, Elbaz et al.\ 1994); fitting
it as a free parameter gave consistent values.  Temperatures for the radial
profiles were fit using XSPEC, and two-dimensional temperature maps were
derived from adaptively smoothed ACIS images in a number of narrow energy
bands as described in Markevitch et al.\ (2000).

\section*{RESULTS AND DISCUSSION}

Figure 1 presents radial temperature profiles for the three distant
clusters, overlaid on the \asca\ projected profiles for A665 and A2163 from
Markevitch (1996). The profiles for A2163 and A2218 show a
radial decline --- for A2163, in good agreement with the \asca\ result from
Markevitch (1996), and for A2218, with that from White (2000). The A665
profile is consistent with a constant temperature, although it does not
cover the complete radial range of the \asca\ data.

However, the A665 temperature map (Fig.\ 2) shows that this cluster is in
fact highly nonisothermal --- there is a spectacular bow shock in front of
what appears from the image to be a cooler cluster core moving from the NW
direction with respect to the gas in the Southern region.

Figure 3 shows a temperature map of A2163. Already the brightness contours
of the ACIS image show that the cluster inner region is probably in a highly
nonrelaxed state. The temperature map confirms that and reveals hot gas
regions coincident with enhancements in the gas density that most plausibly
are shocks or streams of shock-heated gas. The central region is cooler in
projection, perhaps because the shocks have not penetrated the dense
subcluster cores as they fell to the center, similarly to what we see in
A2142 and A3667 (Markevitch et al.\ 2000; Vikhlinin et al.\ 2000).
Temperature maps for A2163 and A665 and their relation to the radio data
will be discussed in detail in the forthcoming paper.

Figure 4 shows a detailed temperature map for A754. On a large scale, it is
in good agreement with earlier results from Henry \& Briel (1995) and
Henriksen \& Markevitch (1996). The latter's \asca\ map had only $3\times 3$
pixels in the area shown in Fig.\ 4. Large hot area South and Southwest of
center and smaller hot regions elsewhere most plausibly result from shock
heating. There is a curious low-temperature region at the tip of the
elongated dense gas body; it does not coincide with either the gas density
peak or one of the two cD galaxies in this cluster. A close examination of
the optical image reveals a possible small galaxy group at this position,
and a simple estimate shows that the elongated body may in fact be a
projection of two more round subclusters of different temperatures, making
A754 a three-body merger. A more detailed discussion will be presented in a
forthcoming paper.

\acknowledgements

The results presented here are made possible by the successful effort of the
entire \chandra\ team to launch and operate the observatory. We thank Bill
Forman, Christine Jones, Dan Harris, Larry David, Paul Nulsen, Hank
Donnelly, Dong-Woo Kim and others for useful discussions.

\begin{figure*}[tb]
\pspicture(0,1.4)(18.5,12)

\rput[tl]{0}(-0.2,12){\epsfxsize=9.0cm \epsfclipon
\epsffile[17 155 515 610]{A665_map.ps_b}}

\rput[tl]{0}(1.5,11.7){\psframebox*{\white \rule{4cm}{5mm}}}

\rput[tl]{0}(0,3.5){
\begin{minipage}{8.75cm}
\small\parindent=3.5mm
{\sc Fig.}~2.---Temperature map of A665. Contours show \mbox{ACIS-I} 0.5--4
keV image (point sources removed), while colors show projected gas
temperature. There is a hot region ``in front'' of the cluster core and a
distinct trace of cooler gas from North-West towards the center, generally
following the gas density elongation.
\par
\end{minipage}
}

\rput[tl]{0}(9.6,12){\epsfxsize=9.0cm \epsfclipon
\epsffile[17 155 515 610]{A2163_map.ps_b}}

\rput[tl]{0}(11.3,11.7){\psframebox*{\white \rule{4cm}{5mm}}}

\rput[tl]{0}(9.8,3.5){
\begin{minipage}{8.75cm}
\small\parindent=3.5mm
{\sc Fig.}~3.---Temperature map of A2163. Contours show ACIS-I 0.8--4.5 keV
image (point sources removed), while colors show gas temperature (areas with
high statistical errors are removed). The image shows a lot of structure,
and the temperature structure generally follows that in the gas density.
\par
\end{minipage}
}

\endpspicture
\end{figure*}

\begin{figure*}[b]
\pspicture(0,0)(11.,12)

\rput[tl]{0}(2.8,12){\epsfxsize=12cm \epsfclipon
\epsffile{A754_map.ps_b}}

\rput[tl]{0}(3,1.4){
\begin{minipage}{11cm}
\small\parindent=3.5mm
{\sc Fig.}~4.---Temperature map of A754. Contours show ACIS-I 0.8--6 keV
image (point sources removed), while colors show gas temperature. The cold spot
does not coincide with either the gas density peak or one of the cD
galaxies. 
\par
\end{minipage}
}
\endpspicture
\end{figure*}

\end{document}